\documentclass[prb,twocolumn,showpacs]{revtex4}

\usepackage[dvips]{graphicx}
\usepackage{amsmath,amssymb}
\usepackage{bm}

\newcommand{\beq}{\begin{equation}}
\newcommand{\eeq}{\end{equation}}

\newcommand{\beqn}{\begin{eqnarray}}
\newcommand{\eeqn}{\end{eqnarray}}
\newcommand{\mbold}[1]{\mbox{\boldmath $ #1 $}}

\begin{document}

\title{Doping control of realization of an extended Nagaoka ferromagnetic state from the Mott state}

\author{Hiroaki Onishi}
\email{onishi.hiroaki@jaea.go.jp}
\affiliation{%
Advanced Science Research Center, Japan Atomic Energy Agency,
Tokai, Ibaraki 319-1195, Japan
}
\author{Seiji Miyashita}
\email{miya@spin.phys.s.u-tokyo.ac.jp}
\affiliation{%
Department of Physics, Graduate School of Science,
The University of Tokyo, 7-3-1 Hongo, Bunkyo-Ku, Tokyo 113-8656, Japan
}
\affiliation{%
CREST, JST, 4-1-8 Honcho Kawaguchi, Saitama 332-0012, Japan
}

\date{January 13, 2015}

\begin{abstract}
Inspired by the Nagaoka ferromagnetism,
we propose an itinerant model to study the transition
between the Mott singlet state and a ferromagnetic state
by emulating a doping process in finite lattices.
In the Nagaoka ferromagnetism,
the total spin of the system takes the maximum value
when an electron is removed from the half-filled system.
To incorporate a procedure of the electron removal,
our model contains extra sites as a reservoir of electrons,
and the chemical potential of the reservoir controls the distribution of electrons.
As a function of the chemical potential,
the system exhibits ground-state phase transitions
among various values of the total spin,
including a saturated ferromagnetic state due to the Nagaoka mechanism at finite hole density.
We discuss the nature of the ferromagnetism by measuring various physical quantities,
such as the distribution of electrons, the spin correlation functions,
the magnetization process in the magnetic field, 
and also the entanglement entropy.
\end{abstract}



\maketitle

\section{INTRODUCTION}

The itinerant ferromagnetism is inherently a quantum phenomenon
in which the electron correlation is essential.
Microscopic origin for the itinerant ferromagnetism has been studied
in the framework of
the Hubbard model.\cite{Hubbard1963,Kanamori1963,Gutzwiller1963}
A conventional mean-field treatment leads to the Stoner criterion
for the occurrence of band ferromagnetism.\cite{Stoner1946}
It tells us that
the ferromagnetism occurs
if the Coulomb repulsion and/or
the density of states at the Fermi level are large enough.
However, such a mean-field treatment
overestimates the stability of the ferromagnetism
and it is not adequate for the effect of the electron correlation.
In fact, according to the multiple scattering theory,\cite{Kanamori1963}
the Coulomb repulsion is revised by a renormalized one in the Stoner criterion,
which improves the stability condition.
On the other hand,
the occurrence of the ferromagnetism in the Hubbard model has been proven rigorously
in some limiting conditions.\cite{Nagaoka1966,Mielke1993,Tasaki1998}
The Nagaoka ferromagnetism is a well-known rigorous result,\cite{Nagaoka1966}
indicating that systems have a saturated ferromagnetic ground state
when there is one hole added to the half-filling and the Coulomb repulsion is infinitely large
on appropriate lattices that satisfy the so-called connectivity condition.

Note that the introduction of one hole
corresponds to an infinitesimal hole doping in the thermodynamic limit.
In order to explore the ferromagnetism in more realistic conditions,
much effort has been made to know
how the ferromagnetic phase extends
in the case with finite holes
and finite Coulomb repulsion.\cite{Takahashi1982,Doucot1989,Riera1989,Fang1989,
Shastry1990,Doucot1990,Putikka1992,Becca2001,Carleo2011,Liu2012,
Liang1995,Kohno1997,Watanabe1997,Watanabe1999,Kollar1996}
In a square lattice, for instance,
many authors have tried to clarify
whether the saturated ferromagnetic state survives
over a finite range of hole density,\cite{Shastry1990,Doucot1990,Putikka1992,
Becca2001,Carleo2011,Liu2012}
and most results are supportive for the ferromagnetism at finite hole densities.
In a two-leg ladder,
numerical results have shown that
the saturated ferromagnetic ground state keeps stable
up to a critical hole density.\cite{Liu2012,Liang1995,Kohno1997}
The critical hole density is insensitive to the ladder width
for wider four-leg and six-leg ladders,
suggesting that the ferromagnetism in the two-leg ladder smoothly connects
with that in two dimensions as the ladder width increases.\cite{Liu2012,Liang1995}
Regarding the experimental observation of the Nagaoka ferromagnetism,
it has been proposed to use cold-atom optical lattice systems,
since the Hubbard model is realizable in a clean environment
with high tunability and controllability.\cite{Stecher2010,Okumura2011}

\begin{figure}[t] 
\begin{center}
\includegraphics[clip,width=4cm]{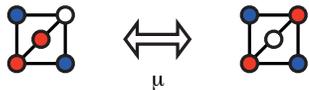}
\end{center}
\caption{(Color online)
A lattice with five sites,
which is used as a unit structure to construct an extended lattice.
Here we put four electrons.
Solid circles are occupied by an electron,
while open circles are vacant.
The electron occupation is controlled by the chemical potential at the center site $\mu$.
}
\label{Fig_lattice5}
\end{figure}

In sharp contrast to the Nagaoka ferromagnetism,
the ground state at half-filling
is a Mott state with zero total spin
in a bipartite lattice with an equal number of sites in each sublattice.\cite{Lieb-Mattis}
Namely,
with the change of the number of electrons by one,
we can see a drastic change
between the Mott state with zero total spin
and the ferromagnetic state with the maximum total spin.
Since the number of electrons is a conserved quantity in the Hubbard model,
we usually study these states independently
by changing the number of electrons one by one.
To shed light on the quantum mechanical transition between the two states,
we have introduced a quantum mechanical procedure
to remove an electron from the system,
considering a model with a four-site plaquette and an extra site,
as depicted in Fig.~\ref{Fig_lattice5}.~\cite{PTPmiya}
There
the electron occupation is controlled
through a chemical potential at the extra site.
The extra site can be regarded as a particle reservoir
for a part of the system without the extra site.
We have also discussed types of itinerant ferromagnetism
for particles with $S>1/2$,
which could be realized in optical lattices with laser-cooled atoms.\cite{PRBmiya}

In this paper
we study a mechanism for the control of the magnetic property
through a local chemical potential
in systems which are made by a five-site unit structure in Fig.~\ref{Fig_lattice5}.
We investigate the ground-state properties
by numerical methods
such as Lanczos diagonalization
and density-matrix renormalization group (DMRG).\cite{White1992,White1993}
As the chemical potential is varied,
we observe ground-state transitions among various values of the total spin
due to the change of the distribution of electrons.
In particular,
we find a saturated ferromagnetic state in a broad region of the chemical potential
irrespective of the system size,
suggesting that the ferromagnetism is realized
due to the present mechanism in the thermodynamic limit.

The organization of the paper is as follows.
In Sec.~II
we explain our basic idea of a mechanism to switch the ground state
between the Mott state and the Nagaoka ferromagnetic state,
based on the case of a small five-site lattice.\cite{PTPmiya}
In Sec.~III
we introduce an extended model of larger sites.
We present numerical results to demonstrate that
a saturated ferromagnetic state occurs.
In Sec.~IV
we discuss the characteristics of phases
from a viewpoint of correlations.
Section V is devoted to summary and discussion.

\section{ROUTE FROM MOTT STATE TO FERROMAGNETIC STATE}

\subsection{Ferromagnetism in itinerant electron system: Nagaoka ferromagnetism}

The Hubbard model is one of the simplest models for itinerant electron systems.
It is composed of the electron hopping term and the on-site Coulomb repulsion term,
described by
\begin{equation}
  {\cal H}=
  -t \sum_{\langle ij \rangle,\sigma} (c_{i\sigma}^{\dagger}c_{j\sigma}+{\rm h.c.})
  +U \sum_{i} n_{i\uparrow}n_{i\downarrow},
\label{Eq_hubbard}
\end{equation}
where
$c_{i\sigma}$ is an annihilation operator of an electron
with spin $\sigma$ $(=\uparrow,\downarrow)$ at site $i$,
$n_{i\sigma}=c_{i\sigma}^{\dagger}c_{i\sigma}$,
$t$ is the hopping amplitude,
and $U$ is the on-site Coulomb repulsion.
We set $t=1$ and take it as the energy unit.
Because of the SU(2) symmetry,
the ground state is characterized by the total spin $S_{\rm tot}$,
given by
\begin{equation}
  S_{\rm tot}(S_{\rm tot}+1) = \langle \mbold{S}_{\rm tot}^{2} \rangle,
\label{Eq_Stot}
\end{equation}
where
$\mbold{S}_{\rm tot}=\sum_{i}\mbold{S}_{i}$
and $\langle \cdots \rangle$ denotes the expectation value in the ground state.
Note that spin-1/2 operators are described by
the electron creation and annihilation operators
and the Pauli matrices $\bm{\sigma}$ as
\begin{equation}
  \mbold{S}_{i}=
  \frac{1}{2}\sum_{\sigma,\sigma'}
  c_{i\sigma}^{\dag}
  \mbox{\boldmath $\sigma$}_{\sigma\sigma'}
  c_{i\sigma'}.
\end{equation}
The total magnetization, defined by
\begin{equation}
  M=\sum_{i} \langle S_{i}^{z} \rangle,
\end{equation}
is also a conserved quantity.
In the following,
we analyze the ground state with $M=0$ unless otherwise specified.

The ground state at half-filling, i.e., one electron per site
($N_{\rm e}=N$,
where $N$ is the number of sites and $N_{\rm e}$ is the number of electrons),
is a Mott state with zero total spin in a bipartite lattice
with equal number of sites in each sublattice.
In contrast,
if we remove one electron from the half-filling ($N_{\rm e}=N-1$),
the ground state is a saturated ferromagnetic state with the maximum total spin,
assuming that $U$ is sufficiently large
and the lattice satisfies the connectivity condition,
which is well known as the Nagaoka ferromagnetism.~\cite{Nagaoka1966}
Here we note that the hopping coefficient $-t$ is negative,
and in order to ensure the symmetric ground state,
the lattice should have a bipartite structure,
in which we can change the sign of the hopping amplitude
by a gauge transformation
$c_{i\sigma} \rightarrow -c_{i\sigma}$.

In the present study, as we will explain in the next subsection (Sec.~II~B),
instead of removing one electron from the half-filled system,
we consider a kind of particle bath
to control the number of electrons in a part of the system effectively.
That is, the system is composed of a subsystem and the particle bath.
We expect a kind of Nagaoka ferromagnetism
when the number of electrons in the subsystem is reduced from
the half-filled case of the subsystem.
We call the present mechanism ``extended Nagaoka ferromagnetism.''

\subsection{Control of total spin by mechanism of Nagaoka ferromagnetism}

We have proposed a possible mechanism to switch the ground state
between the Mott state and the Nagaoka ferromagnetic state.~\cite{PTPmiya}
Here we explain its basic idea for the completeness of the paper,
and make a few remarks relevant for the present study.

Since the number of electrons is a conserved quantity in the Hubbard model,
it is difficult to describe a process to remove an electron
from the system in a Hamiltonian.
In order to describe a procedure for the electron removal,
we prepare an extra site to which an electron can escape.
In Fig.~\ref{Fig_lattice5}
we present an example of such a lattice with five sites where we put four electrons.
The lattice is composed of a four-site ring and a center site.
We call the four-site ring ``subsystem.''
The Hamiltonian is explicitly given by
\begin{eqnarray}
  {\cal H}
  &=&
  -t \sum_{\langle ij \rangle,\sigma} (c_{i\sigma}^{\dagger}c_{j\sigma}+{\rm h.c.})
  +U \sum_{i} n_{i\uparrow}n_{i\downarrow}
\nonumber \\
  &&
  +\mu (n_{5\uparrow}+n_{5\downarrow}),
\label{Eq_hammu4}
\end{eqnarray}
where $\langle ij \rangle$ denotes pairs of sites
connected by a solid line in Fig.~\ref{Fig_lattice5},
the center site is named ``5,''
and $\mu$ is the on-site energy,
which we call ``chemical potential.''
Note that the Coulomb repulsion is active in all the sites
including the subsystem and the center site.
$S_{\rm tot}$ and $M$ are conserved quantities
in the same way as the Hubbard model (\ref{Eq_hubbard}).

We note that in the previous paper~\cite{PTPmiya}
we defined the chemical potential with the opposite minus sign.
In this paper
we will introduce an extended lattice composed of a subsystem and center sites
in the next section.
We regard the center sites as a reservoir of electrons,
and the present sign is more appropriate in its physical meaning.
That is, if the chemical potential is large,
electrons tend to move out from the center sites.
The chemical potential represents the electron affinity at the center site.

\begin{figure}[t] 
\begin{center}
\includegraphics[clip,width=8cm]{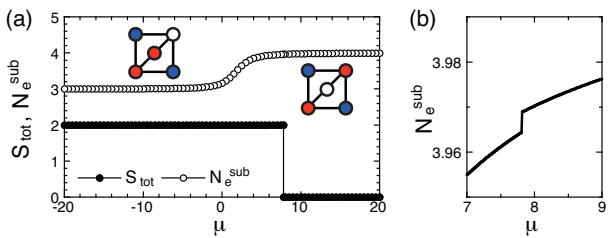}
\end{center}
\caption{(Color online)
(a) The $\mu$ dependence of the total spin $S_{\rm tot}$
and the number of electrons in the subsystem $N_{\rm e}^{\rm sub}$
for the system with five sites and four electrons at $U=1000$.
(b) $N_{\rm e}^{\rm sub}$ around $\mu=8$ in a magnified scale,
where we clearly see a jump.
}
\label{Fig_stot-nsub_n5u2d2_u1000}
\end{figure}

In Fig.~\ref{Fig_stot-nsub_n5u2d2_u1000}
we plot the total spin $S_{\rm tot}$
and the number of electrons in the subsystem $N_{\rm e}^{\rm sub}$, given by
\begin{equation}
  N_{\rm e}^{\rm sub} =
  \sum_{i\in {\rm subsystem},\sigma} \langle n_{i\sigma} \rangle,
\end{equation}
where the summation of $i$ is taken over the sites of the subsystem,
as a function of $\mu$ at $U=1000$.
We clearly see a ground-state transition at around $\mu=8$,
where $S_{\rm tot}$ exhibits a jump.
We also find a small jump of $N_{\rm e}^{\rm sub}$,
as depicted in the magnified figure.
At the transition point,
$N_{\rm e}^{\rm sub}$ is about $3.97$.
We envisage that this situation would correspond to a hole doping into the subsystem
to result in a change of the magnetic property.

For large $\mu$,
electrons are repelled from the center site and stay in the subsystem.
The subsystem is half-filled
and the ground state is essentially in the Mott state
with antiferromagnetic correlations.
For negative $\mu$,
electrons are attracted onto the center site,
so that an electron is removed from the subsystem.
Then the total spin takes the maximum value
$S_{\rm tot}=S_{\rm tot}^{\rm max}=2$,
indicating a fully symmetrized state.
We should remark that the Nagaoka ferromagnetism in the subsystem
leads to the total spin of the subsystem of $3/2$.
Note that in the present case,
the total spin of the subsystem $S_{\rm sub}$ and that of the center site $S_{\rm c}$
are not conserved quantities,
but they are still approximately represented by $S_{\rm sub}=3/2$ and $S_{\rm c}=1/2$,
where we define the total spin of a part of the system as
\begin{equation}
  S_{\rm sub}(S_{\rm sub}+1) = \langle\mbold{S}_{\rm sub}^{2}\rangle,
\end{equation}
\begin{equation}
  S_{\rm c}(S_{\rm c}+1) = \langle\mbold{S}_{\rm c}^{2}\rangle,
\end{equation}
with
$\mbold{S}_{\rm sub}=\sum_{i\in {\rm subsystem}}\mbold{S}_{i}$
and
$\mbold{S}_{\rm c}=\sum_{i\in {\rm center}}\mbold{S}_{i}$.
We observe that
the total spin of the whole system is given by
the combination of those of the subsystem and the center site as
$S_{\rm tot}=3/2+1/2=2$.
Although this could be, in principle,
$S_{\rm tot}=3/2-1/2=1$,
the symmetrized state is realized in the present case.
In this way, we can control the total spin by the chemical potential in a local site.

If we further decrease $\mu$ down to  $\mu \lesssim -U$,
two electrons are trapped at the center site,
suggesting that we could possibly change $N_{\rm e}^{\rm sub}$ in a wider range,
including cases of the doubly occupied center site.
In fact, for huge negative $\mu$ around $\mu=-U$,
it is observed that the center site traps two electrons.
However, it turns out that $S_{\rm tot}$ is simply zero without exhibiting any magnetic states.
This fact is similarly found in an extended lattice
introduced in the following section.
In the present paper,
we focus on the case of $|\mu| \ll U$,
since we find a complete ferromagnetic state only for $|\mu| \ll U$
in an extended lattice.

\section{Extended lattice}

\begin{figure}[t] 
\begin{center}
\includegraphics[clip,width=6cm]{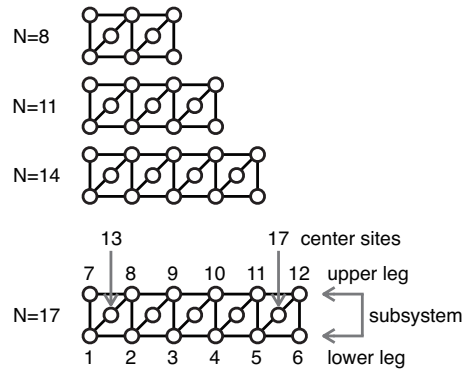}
\end{center}
\caption{
Extended lattices with 8, 11, 14, and 17 sites.
The site numbering is presented for $N=17$ as an example.
The lattice sites are labeled
in the order of the sites in the lower leg in the subsystem,
those in the upper leg in the subsystem,
and those in the center sites.
}
\label{Fig_extended-lattice}
\end{figure}

Here we extend the five-site lattice depicted in Fig.~\ref{Fig_lattice5} to larger system sizes
by simply repeating it as a unit in one direction,
as shown in Fig.~\ref{Fig_extended-lattice}.
The extended lattice is composed of a subsystem in ladder shape and center sites.
For instance, the lattice of 11 sites consists of three units,
and the system has a subsystem of an eight-site ladder and three center sites.
The number of sites $N$ is given by
the sum of that in the subsystem $N^{\rm sub}$ and that in the center sites $N^{\rm c}$
as $N=N^{\rm sub}+N^{\rm c}$.
We consider the same type of Hamiltonian as Eq.~(\ref{Eq_hammu4}),
\begin{eqnarray}
  {\cal H}
  &=&
  -t \sum_{\langle ij \rangle,\sigma} (c_{i\sigma}^{\dagger}c_{j\sigma}+{\rm h.c.})
  +U \sum_{i} n_{i\uparrow}n_{i\downarrow}
\nonumber \\
  &&
  +\mu \sum_{i \in {\rm center}} (n_{i\uparrow}+n_{i\downarrow}),
\label{Eq_hammu}
\end{eqnarray}
where the chemical potential $\mu$ is effective for the center sites.
We use open boundary conditions.

Let us consider the system with electrons which fill the subsystem,
i.e., $N_{\rm e}=N^{\rm sub}$.
For instance, in the lattice of 11 sites,
we put eight electrons to fill the outside eight-site ladder.
This is a natural extension of the five-site case in Sec.~II,
since the center sites are empty and the subsystem is half-filled for large $\mu$,
while electrons turn to occupy the center sites
as $\mu$ decreases.

We mention that the present system is also regarded as a periodic Anderson lattice,
which is a typical model for heavy-electron systems,
assuming that
the subsystem represents conduction-electron sites
and the center sites correspond to $f$-electron sites.
Here $\mu$ plays a role of a local $f$-electron level.
Regarding the hybridization,
due to the lattice structure in Fig.~\ref{Fig_extended-lattice},
each $f$-electron site is connected to two conduction-electron sites,
indicating a multiband system.
Moreover,
when the $f$-electron sites are singly occupied and they are considered as localized spins,
the system is equivalent to a Kondo lattice.
The ferromagnetism in periodic Anderson and Kondo lattice models has been studied extensively.
\cite{Rice1985,Rice1986,Guerrero1996,Batista2002,Batista2003,Tsunetsugu1997}

We investigate the magnetic properties by making use of numerical techniques.
We examine the $\mu$ dependence of various physical quantities with $U=1000$ fixed.
The $U$ dependence is also discussed in Sec.~III~B.
We use the Lanczos diagonalization method for small clusters up to $N=17$
to obtain numerical results with relatively small computational costs.
For larger lattices,
we also perform extensive DMRG calculations
to grasp the ground-state properties in the thermodynamic limit.
We note that we adopt open boundary conditions for Lanczos and DMRG calculations
for consistency.

\subsection{Extended Nagaoka ferromagnetic state}

\subsubsection{Lanczos results}

Before going into the discussion of Lanczos results,
let us make a few comments on technical details.
Because of the SU(2) symmetry,
we usually determine the total spin from the ground-state degeneracy.
That is, by comparing the ground-state energies
of different values of the total magnetization $M$,
if the ground states with $\vert M \vert \le S$ are degenerate,
the total spin is estimated to be $S$.
However,
this is a hard task in the present model.
Since the model involves charge and spin degrees of freedom,
many nearly degenerate low-energy states appear
because of a subtle balance of multiple degrees of freedom.
In such a case,
it is difficult to determine the total spin
by comparing the ground-state energies numerically.
Instead, we obtain the ground-state wave function with $M=0$
and evaluate $S_{\rm tot}$ by using Eq.~(\ref{Eq_Stot}).
In addition,
we need to perform a large number of Lanczos iteration steps
to obtain an accurate ground-state wave function
which gives an integer value of $S_{\rm tot}$.
Typically,
several thousand steps are required for the good convergence
near transition points
and in the negative $\mu$ region.

\begin{figure}[t] 
\begin{center}
\includegraphics[clip,width=6.5cm]{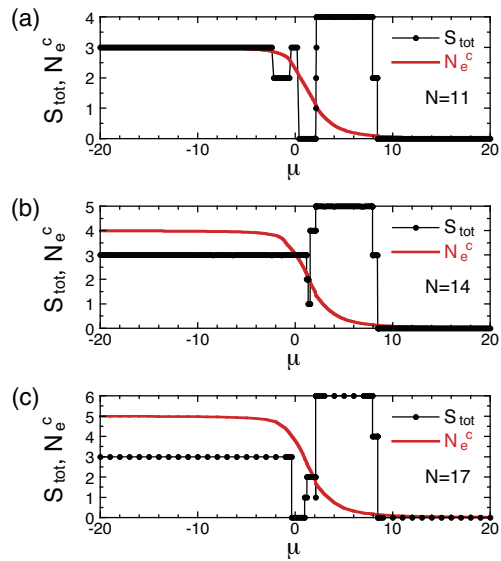}
\end{center}
\caption{(Color online)
The total spin $S_{\rm tot}$
and the number of electrons in the center sites $N_{\rm e}^{\rm c}$
for several system sizes:
(a) $(N,N_{\rm e})=(11,8)$,
(b) $(14,10)$, and
(c) $(17,12)$.
Here we set $U=1000$.
}
\label{Fig_stot-nc-U1000}
\end{figure} 

In Fig.~\ref{Fig_stot-nc-U1000}
we show the $\mu$ dependence of the total spin $S_{\rm tot}$
and the number of electrons in the center sites $N_{\rm e}^{\rm c}$,
given by
\begin{equation}
  N_{\rm e}^{\rm c} =
  \sum_{i\in {\rm center},\sigma} \langle n_{i\sigma} \rangle,
\end{equation}
for $N=11$, $14$, and $17$.
Note that $N_{\rm e}=N_{\rm e}^{\rm sub}+N_{\rm e}^{\rm c}$ by definition.
For large $\mu$, the center sites are vacant
and the subsystem is half-filled,
so that the Mott state with $S_{\rm tot}=0$ is realized.
With decreasing $\mu$,
$N_{\rm e}^{\rm c}$ gradually increases,
since electrons come to the center sites.
The same amount of holes is introduced into the subsystem.
In such a situation
we expect a ground-state change from the Mott state to a ferromagnetic state
in a similar way to the five-site model.
Indeed,
we observe that $S_{\rm tot}$ jumps from zero to $N_{\rm e}/2-2$
and stays there in a short period around $\mu=8$,
and then it increases up to the maximum value
$N_{\rm e}/2$.
With further decreasing $\mu$,
$S_{\rm tot}$ is reduced from $N_{\rm e}/2$.
Sudden jumps of $S_{\rm tot}$ signal transitions of first order.
We note that
the complete ferromagnetic state is found
in $2 \lesssim \mu \lesssim 8$
similarly for $N=11$, $14$, and $17$,
implying that the complete ferromagnetic state is realized
without significant finite-size effects.
Note also that
the complete ferromagnetic state appears
in a region where the amount of holes doped into the subsystem is moderately small.

Here let us discuss the underlying mechanism of the complete ferromagnetic state
from the viewpoint of hole doping into the subsystem.
We note that
the subsystem is equivalent to a two-leg ladder,
eliminating the center sites from the whole system.
Thus we refer to
the ground state of the two-leg ladder Hubbard model
as a function of the hole doping rate.~\cite{Liu2012,Liang1995,Kohno1997}
When we have one electron less than half-filling,
the Nagaoka ferromagnetism takes place.
Even when we add holes,
the ground state remains a ferromagnetic state due to the Nagaoka mechanism
in some range of the hole density.
When the hole density exceeds a critical point,
the ground state changes to a partially spin-polarized state,
and eventually becomes a spin-singlet state.
In the present model we expect the same behavior for the subsystem
when $\mu$ varies.
That is,
the subsystem exhibits a ferromagnetic state in a range of $\mu$,
and it is destabilized with the decrease of $\mu$ due to hole doping into the subsystem.
As a result,
the complete ferromagnetic state in the whole system is broken down.
Thus, the complete ferromagnetic state is attributed to
the Nagaoka ferromagnetism at finite hole density in the subsystem.

We mention that in the region near $\mu=0$,
$S_{\rm tot}$ shows a complicated dependence.
When $\mu$ is decreased further,
all the center sites are singly occupied,
corresponding to the Kondo lattice regime.
In that region
we do not find the complete ferromagnetic state with $S_{\rm tot}=N_{\rm e}/2$,
but observe that $S_{\rm tot}=3$
for $N=11$, $14$, and $17$.
We would expect that
$S_{\rm tot}=3$ even for larger system sizes,
but actually this is not the case,
as we will discuss based on DMRG results later.

\begin{figure}[t] 
\begin{center}
\includegraphics[clip,width=6.5cm]{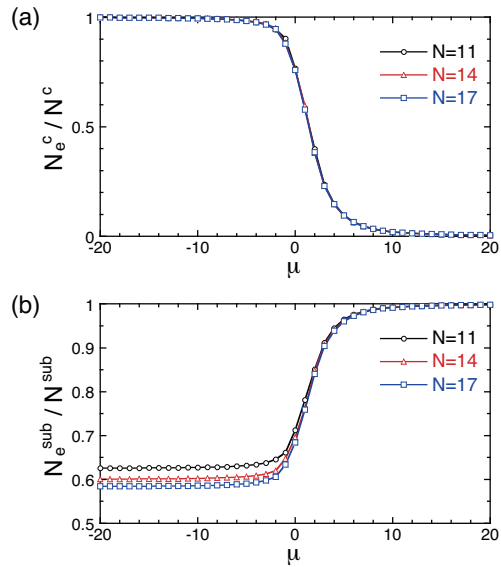}
\end{center}
\caption{(Color online)
(a) The electron density in the center sites $N_{\rm e}^{\rm c}/N^{\rm c}$,
and
(b) that in the subsystem $N_{\rm e}^{\rm sub}/N^{\rm sub}$,
for several system sizes at $U=1000$.
}
\label{Fig_ncall}
\end{figure}

In Fig.~\ref{Fig_ncall}(a)
we show the $\mu$ dependence of the electron density in the center sites
$N_{\rm e}^{\rm c}/N^{\rm c}$
for $N=11$, $14$, and $17$ in the same plot.
We clearly observe that
with decreasing $\mu$,
$N_{\rm e}^{\rm c}/N^{\rm c}$ gradually increases from zero to unity,
and its $\mu$ dependence is independent of the system size.
On the other hand,
as shown in Fig.~\ref{Fig_ncall}(b),
the electron density in the subsystem $N_{\rm e}^{\rm sub}/N^{\rm sub}$
decreases from unity to a constant value that depends on $N$,
\begin{equation}
  \frac{N_{\rm e}^{\rm sub}}{N^{\rm sub}}
  = \frac{N^{\rm sub}-N^{\rm c}}{N^{\rm sub}},
\label{Eq_nesub}
\end{equation}
and it becomes $0.5$ in the large $N$ limit.
Here electrons escape from the subsystem to the center sites.
In other words, holes are doped into the subsystem.
The doping rate is controlled by $\mu$.
In this situation
we can regard the center sites as a reservoir of electrons,
i.e., a kind of particle bath.

\subsubsection{DMRG results}

As for the efficiency of the present DMRG calculations,
we mention that the computational cost
highly depends on the value of $\mu$.
At large $\mu$,
the center sites are vacant and the subsystem is half-filled,
so that charge degrees of freedom are frozen out
and only spin degrees of freedom are relevant in the strong-coupling regime.
In such a case we can easily obtain the ground state in high precision.
However,
with decreasing $\mu$,
charge degrees of freedom should also become relevant,
since electrons turn to move around the whole system.
This indicates that we need to keep a large number of DMRG states
to describe the ground state.
For instance,
in the region near $\mu=0$,
even if we keep 1000 states,
in which the truncation error is estimated to be around $10^{-6}$,
it is still difficult to obtain the true ground state,
and we have incorrect results such as a non-integer value of $S_{\rm tot}$.

\begin{figure}[t] 
\begin{center}
\includegraphics[clip,width=6.5cm]{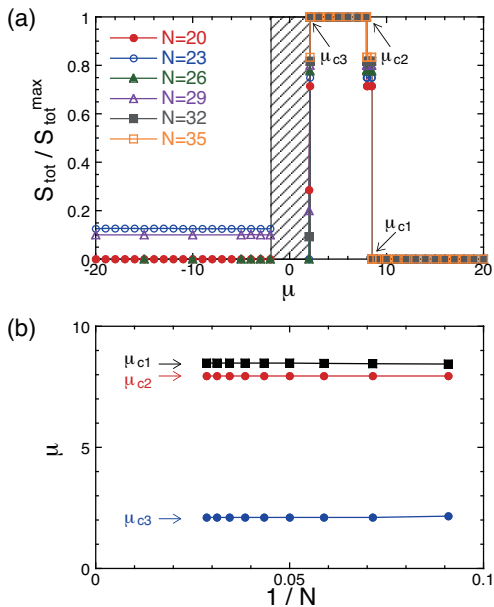}
\end{center}
\caption{(Color online)
DMRG results with large system sizes up to $N=35$ at $U=1000$.
(a) The total spin $S_{\rm tot}$ normalized by
the maximum value $S_{\rm tot}^{\rm max}=N_{\rm e}/2$,
i.e., $0 \le S_{\rm tot}/S_{\rm tot}^{\rm max} \le 1$ regardless of the system size.
In the shaded region near $\mu=0$
we cannot obtain well-converged data by DMRG simulations even if we keep 1000 states.
(b) The size dependence of the transition points $\mu_{\rm c}$'s
denoted by arrows in (a).
Here we also plot Lanczos results for $N=11$, $14$, and $17$ together.
}
\label{Fig_DMRG}
\end{figure} 

Figure~\ref{Fig_DMRG}(a) presents DMRG results of $S_{\rm tot}$
for various values of the system size $N$.
We find that with decreasing $\mu$,
$S_{\rm tot}$ changes from zero to $N_{\rm e}/2-2$
and then it takes the complete ferromagnetic value $N_{\rm e}/2$
in a broad region $2 \lesssim \mu \lesssim 8$
in the same way as Lanczos results in Fig.~\ref{Fig_stot-nc-U1000}.
This tendency is found for all the system sizes we have analyzed.
In Fig.~\ref{Fig_DMRG}(b)
we show the size dependence of critical points of $\mu$
where $S_{\rm tot}$ changes:
$\mu_{\rm c1}$ for the change between $S_{\rm tot}=0$ and $N_{\rm e}/2-2$;
$\mu_{\rm c2}$ for the change between $S_{\rm tot}=N_{\rm e}/2-2$ and $N_{\rm e}/2$;
and
$\mu_{\rm c3}$ for the change between $S_{\rm tot}=N_{\rm e}/2$ and a lower value.
We clearly find that they saturate well at large values of $N$
without any significant size dependence,
and thus we believe that these critical values exist in the thermodynamic limit.
Moreover, the $\mu$ dependencies of $N_{\rm e}^{\rm c}$
of the systems with $17<N\leq 35$ collapse
as we saw for $N\leq 17$ in Fig.~\ref{Fig_ncall} (not shown).

On the other hand,
we do not see saturated behavior of $S_{\rm tot}$
as a function of $N$ for negative $\mu$.
In fact,
$S_{\rm tot}=3$ for $N \le 17$,
but $S_{\rm tot}=0$ for $N=20$ and $26$,
and $S_{\rm tot}=1$ for $N=23$ and $29$.
This complication would be probably due to the fact that
the electron density in the subsystem varies with the system size
as Eq.~(\ref{Eq_nesub}).
Moreover, we point out that
even if we keep 1000 states,
the DMRG does not give fairly well-converged data
in the region near $\mu=0$,
denoted by the shaded region in Fig.~\ref{Fig_DMRG}(a).
Note that the Lanczos results for small clusters also show
the complicated dependence near $\mu=0$.
These regions are interesting to study
for a possible realization of magnetic states other than the complete ferromagnetic state
such as a partially spin-polarized state,
but we leave it for a future issue.

\subsection {$U$ dependence}

\begin{figure}[t] 
\begin{center}
\includegraphics[clip,width=6.5cm]{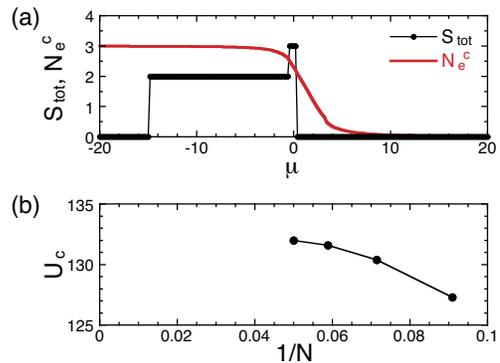}
\end{center}
\caption{(Color online)
(a) The total spin $S_{\rm tot}$
and the number of electrons in the center sites $N_{\rm e}^{\rm c}$
at $U=100$ for $N=11$.
(b) The size dependence of the critical value of $U$
for the appearance of the ferromagnetism $U_{\rm c}$
with $N=11$, $14$, $17$, and $20$.
}
\label{Fig_stot-nc-U100_Uc}
\end{figure}

It should be noted that
the value of $U$ which can cause the comoplete ferromagnetic state
depends on the system size.
In fact,
$U=100$ is not large enough to realize the complete ferromagnetic state for $N \ge 11$,
while $U=100$ was enough to generate it in the system of $N=5$.\cite{PTPmiya}
As a typical example
we depict $S_{\rm tot}$ and $N_{\rm e}^{\rm c}$ for $N=11$ at $U=100$
in Fig.~\ref{Fig_stot-nc-U100_Uc}(a).
There $S_{\rm tot}$ does not reach the maximum value $S_{\rm tot}^{\rm max}=4$
for any values of $\mu$,
although $N_{\rm e}^{\rm c}$ varies with $\mu$
in the same way as compared with the case of $U=1000$ in Fig.~\ref{Fig_stot-nc-U1000}(a).
The critical value of $U$ above which the complete ferromagnetic state appears
is estimated at $U_{\rm c}=127.3$ for $N=11$,
and it further shifts to $U_{\rm c}=130.4$ for $N=14$,
indicating finite-size effects on the location of the phase boundary in the region of small $U$.
In Fig.~\ref{Fig_stot-nc-U100_Uc}(b)
we plot the size dependence of $U_{\rm c}$.
The curve is bent such that the slope becomes gentle as the system size increases,
which suggests a tendency toward convergence.
Note, however, that
the systems studied are still small and
we need calculations with larger sizes for the extrapolation.
In contrast, for large $U$,
we find no significant size dependence at $U=1000$ up to $N=35$,
as shown in Fig.~\ref{Fig_DMRG}(b).
In the present paper
we have used $U=1000$ which is sufficiently large
enough to realize the complete ferromagnetic state.

\begin{figure}[t] 
\begin{center}
\includegraphics[clip,width=6.5cm]{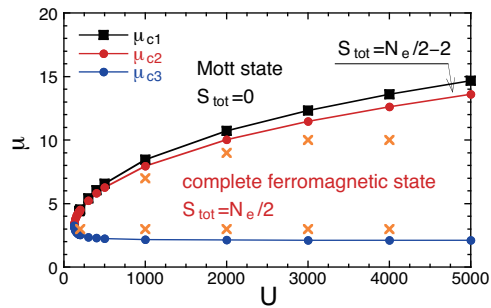}
\end{center}
\caption{(Color online)
The ground-state phase diagram in the coordinate $(U,\mu)$ for $N=11$.
The transition points $\mu_{\rm c}$'s are defined
in the same way as those in Fig.~\ref{Fig_DMRG}.
At several points denoted by crosses,
we have confirmed the realization of the complete ferromagnetic state
by DMRG calculations up to $N=110$.
}
\label{Fig_muc-U_n11ne8}
\end{figure}

In Fig.~\ref{Fig_muc-U_n11ne8}
we present the ground-state phase diagram
in the coordinate $(U,\mu)$ for $N=11$.
The region between $\mu_{\rm c2}$ and $\mu_{\rm c3}$ is
of the complete ferromagnetic state with $S_{\rm tot}=N_{\rm e}/2$.
We clearly see that
the range of $\mu$ of the complete ferromagnetic state shrinks with decreasing $U$,
and it eventually disappears at small $U$,
as we mentioned above.
Here we note again that
finite-size effects should be carefully considered
to discuss the phase diagram in the thermodynamic limit.
In the region of small $U$,
the phase boundary is supposed to be deformed
to reduce the ferromagnetic region,
since the critical point of $U$ for the appearance of the ferromagnetism
shifts toward larger $U$ with increasing the system size.
For large $U$,
numerical results of $11 \le N \le 35$ at $U=1000$ are indicative that
the finite-size correction is small.
In addition,
at several points denoted by crosses in Fig.~\ref{Fig_muc-U_n11ne8},
we have performed DMRG calculations up to $N=110$
and confirmed that the complete ferromagnetic state is realized.
Thus, these points are expected to be included
in the ferromagnetic phase in the thermodynamic limit,
although it is hard to determine the entire phase boundary
on the basis of the present numerical results for small systems.

\subsection{Spin correlation function}

\begin{figure}[t] 
\begin{center}
\includegraphics[clip,width=6.5cm]{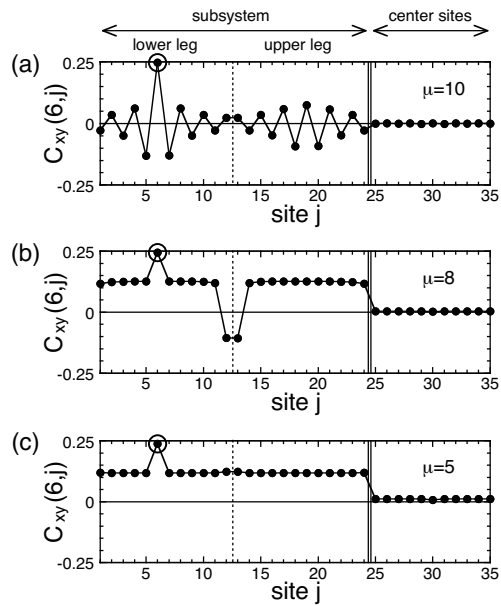}
\end{center}
\caption{
The transverse spin correlation function $C_{xy}(i,j)$
for typical values of $\mu$ at $U=1000$:
(a) $\mu=10$ for the Mott antiferromagnetic state with $S_{\rm tot}=0$;
(b) $\mu=8$ for the partially spin-polarized state with $S_{\rm tot}=N_{\rm e}/2-2$;
and
(c) $\mu=5$ for the complete ferromagnetic state with $S_{\rm tot}=N_{\rm e}/2$,
obtained by DMRG with $N=35$.
The correlation function is measured
with the middle site of the subsystem
(site 6 denoted by an open circle) as starting point.
}
\label{Fig_CSXY}
\end{figure}

In order to clarify the characteristics of magnetic states from a microscopic viewpoint,
it is useful to measure spin correlation functions.
Here we study the ground state in the subspace of $M=0$.
In such a case, when we consider a spin-polarized ground state,
the spin moment lies in the $xy$ plane
and the ferromagnetic correlation develops in the $xy$ plane.
Thus we investigate the transverse spin correlation function,
\begin{equation}
  C_{xy}(i,j) =
  \frac{1}{2}\langle (S_{i}^{x} S_{j}^{x} + S_{i}^{y} S_{j}^{y}) \rangle =
  \frac{1}{4}\langle (S_{i}^{+} S_{j}^{-} + S_{i}^{-} S_{j}^{+}) \rangle.
\end{equation}
In Fig.~\ref{Fig_CSXY}
we show $C_{xy}(i,j)$
measured from the middle of the subsystem
(site $6$ denoted by an open circle),
for typical values of $\mu$.
The plots in the left side of the double line are for the correlation within the subsystem,
and those in the right side denote the correlation between site $6$ in the subsystem and the center sites.
In the subsystem, which has a ladder shape,
the left part of the dotted line represent the lower leg,
while the right part is the upper leg.
The site numbering is given in the bottom of Fig.~\ref{Fig_extended-lattice}.

As shown in Fig.~\ref{Fig_CSXY}(a),
for $\mu=10$,
we find a Ne\'el-type antiferromagnetic correlation
corresponding to the Mott state in the subsystem.
On the other hand, we observe that
the spin correlation between the subsystem and the center sites is almost zero
due to the absence of electrons in the center sites.
At $\mu=8$,
where $S_{\rm tot}=N_{\rm e}/2-2$,
the system is nearly ordered ferromagnetically.
As for the microscopic spin configuration,
we find a ferromagnetic correlation in the subsystem
except for two corner sites,
while the two corner spins align in the opposite direction to the others,
as shown in Fig.~\ref{Fig_CSXY}(b).
At $\mu=5$,
where $S_{\rm tot}=N_{\rm e}/2$,
the system is in the complete ferromagnetic state,
and we find a simple ferromagnetic correlation in the subsystem,
as shown in Fig.~\ref{Fig_CSXY}(c).
We note that the spin correlation between the subsystem and the center sites is still very weak,
since we have only a small number of electrons in the center sites at $\mu=5$.

\begin{figure}[t] 
\begin{center}
\includegraphics[clip,width=6.5cm]{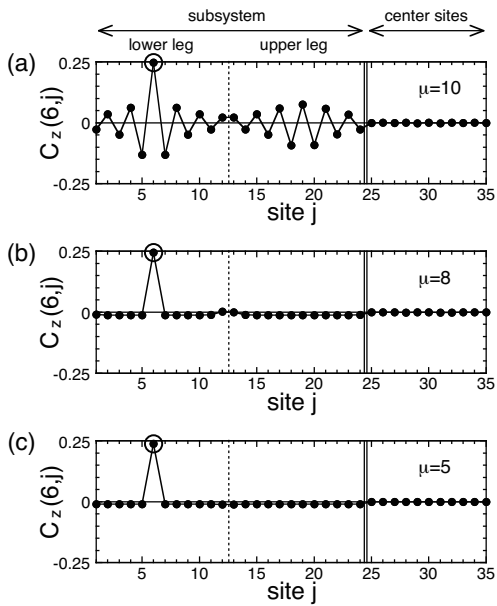}
\end{center}
\caption{
The longitudinal spin correlation function $C_{z}(i,j)$
for typical values of $\mu$ at $U=1000$:
(a) $\mu=10$ for the Mott antiferromagnetic state with $S_{\rm tot}=0$;
(b) $\mu=8$ for the partially spin-polarized state with $S_{\rm tot}=N_{\rm e}/2-2$;
and
(c) $\mu=5$ for the complete ferromagnetic state with $S_{\rm tot}=N_{\rm e}/2$,
obtained by DMRG with $N=35$.
The correlation function is measured
with the middle site of the subsystem
(site 6 denoted by an open circle) as starting point.
}
\label{Fig_CSZ}
\end{figure}

\begin{figure}[t] 
\begin{center}
\includegraphics[clip,width=6cm]{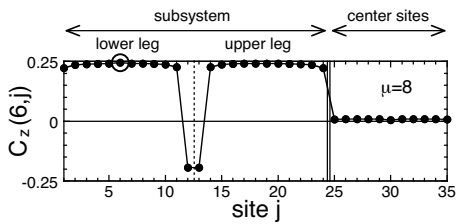}
\end{center}
\caption{
The longitudinal spin correlation function $C_{z}(i,j)$
for the partially spin-polarized state with $S_{\rm tot}=N_{\rm e}/2-2$
in the subspace of $M=N_{\rm e}/2-2$
at $\mu=8$ and $U=1000$,
obtained by DMRG with $N=35$.
The correlation function is measured
with the middle site of the subsystem
(site 6 denoted by an open circle) as starting point.
}
\label{Fig_CSZMMAX}
\end{figure}

Here we investigate the longitudinal spin correlation function,
\begin{equation}
  C_{z}(i,j) = \langle S_{i}^{z} S_{j}^{z} \rangle.
\end{equation}
In Fig.~\ref{Fig_CSZ}(a)
we present $C_{z}(i,j)$ for $\mu=10$.
Since the maximum value of $M$ is zero in the singlet ground state,
the spin correlation is isotropic,
i.e., $C_z(i,j)=C_{xy}(i,j)$.
In contrast,
the ground state is spin-polarized for $\mu=8$ and $5$,
so that we can see anisotropic behavior between $C_{z}(i,j)$ and $C_{xy}(i,j)$,
if we compare them in the subspace of $M=0$.
In Figs.~\ref{Fig_CSZ}(b) and \ref{Fig_CSZ}(c)
we observe that
the longitudinal spin correlation takes a small negative value in the subsystem.
This reflects the sum rule
\begin{equation}
  \langle M^{2} \rangle = \frac{N_{\rm e}}{4}+\sum_{i \ne j}C_{z}(i,j) = 0,
\end{equation}
and we expect that $C_{z}(i,j)\sim -1/N_{\rm e}$.

We notice that the maximum value of $M$ is $N_{\rm e}/2-2$ for $\mu=8$,
suggesting that the spin configuration can be deduced from
$C_{z}(i,j)$ in the subspace of $M=N_{\rm e}/2-2$
instead of $M=0$.
Indeed,
as shown in Fig.~\ref{Fig_CSZMMAX},
we clearly see that the spins at the two corners are antiparallel to the others,
in a similar way to the case of $C_{xy}(i,j)$ in the subspace of $M=0$
shown in Fig.~\ref{Fig_CSXY}(b).

\subsection{Magnetization}

\begin{figure}[t] 
\begin{center}
\includegraphics[clip,width=8cm]{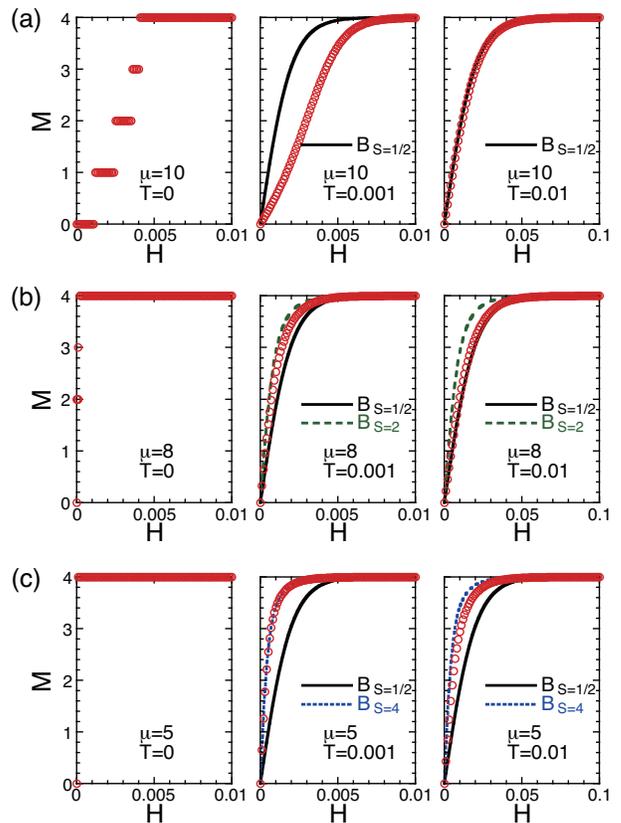}
\end{center}
\caption{(Color online)
The magnetization curve for typical values of $\mu$ at $U=1000$ and several temperatures:
(a) $\mu=10$ for the Mott antiferromagnetic state with $S_{\rm tot}=0$;
(b) $\mu=8$ for the partially spin-polarized state with $S_{\rm tot}=N_{\rm e}/2-2$;
and
(c) $\mu=5$ for the complete ferromagnetic state with $S_{\rm tot}=N_{\rm e}/2$.
For each value of $\mu$
we show magnetization curves at temperatures $T=0$, $0.001$, and $0.01$.
Open circles denote numerical results with $N=11$,
and the Brillouin functions $B_{S}$ for the corresponding temperatures are plotted
by solid or dashed curves.
}
\label{Fig_M}
\end{figure}

We investigate the magnetization curve
as a function of an applied magnetic field $H$,
\begin{equation}
  M(H,T) = \langle M(H) \rangle_{T},
\end{equation}
where $M(H)$ is the magnetization
of model (\ref{Eq_hammu})
with an additional Zeeman term $-H\sum_{i}S_{i}^{z}$,
and $\langle \cdots \rangle_{T}$ is the expectation value at temperature $T$.
Here, for the thermal average,
we need all the eigenvalues and wave functions,
computed by full diagonalization (Householder method).
In Fig.~\ref{Fig_M}
we show the magnetization curve for $N=11$
at typical values of $\mu$ and $T$.
For $\mu=10$,
as shown in the left panel of Fig.~\ref{Fig_M}(a),
the magnetization curve shows a stepwise increase
starting from zero up to the maximum value $4$ at $T=0$,
because $S_{\rm tot}=0$ in the ground state
and there is an energy gap between states of different values of $S_{\rm tot}$
due to the finite system size.
The stepwise structure is rapidly smeared out at finite temperatures,
since the energy gap is rather small.
In the right panel of Fig.~\ref{Fig_M}(a)
we find that
the magnetization curve agrees well with the Brillouin function of $S=1/2$
at $T=0.01$,
indicating that each spin fluctuates independently
due to thermal fluctuations.

Figure~\ref{Fig_M}(b) shows the magnetization curve for $\mu=8$,
where the ground state is a partially spin-polarized state with $S_{\rm tot}=2$.
According to the total spin in the ground state,
the Brillouin function of $S=2$ is realized at low temperature $T=0.001$,
and it changes to that of $S=1/2$ at high temperature $T=0.01$.
In Fig.~\ref{Fig_M}(c)
we can also see a similar behavior for $\mu=5$,
where the ground state is a complete ferromagnetic state with $S_{\rm tot}=4$.
Indeed,
the magnetization curve is represented by the Brillouin function of $S=4$
at $T=0.001$,
and it approaches that of $S=1/2$
as the temperature increases.

\section{CORRELATION BETWEEN SUBSYSTEM AND CENTER SITES}

\subsection{Total spin}

\begin{figure}[t] 
\begin{center}
\includegraphics[clip,width=6.5cm]{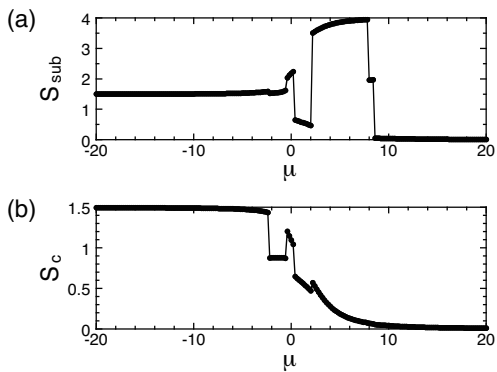}
\end{center}
\caption{
(a) The total spin of the subsystem $S_{\rm sub}$ and
(b) that of the center sites $S_{\rm c}$
at $U=1000$ for $N=11$.
Note that the total spin of the whole system $S_{\rm tot}$ was given
in Fig.~\ref{Fig_stot-nc-U1000}(a).
}
\label{Fig_ssub-sc}
\end{figure}

In Fig.~\ref{Fig_ssub-sc}(a) and \ref{Fig_ssub-sc}(b)
we present the $\mu$ dependence of the total spin of the subsystem $S_{\rm sub}$
and that of the center sites $S_{\rm c}$, respectively,
at $U=1000$.
Comparing with the total spin of the whole system, given in Fig.~\ref{Fig_stot-nc-U1000}(a),
we find that
the total spins of the subsystem and the center sites are correlated positively,
i.e., $S_{\rm tot}=S_{\rm sub}+S_{\rm c}$,
for most values of $\mu$ except for the region near $\mu=0$.
We note that they couple negatively,
i.e., $S_{\rm tot}=|S_{\rm sub}-S_{\rm c}|$, for $0.4 \lesssim \mu \lesssim 2.1$,
and the total spin takes an intermediate value for $-2.2 \lesssim \mu \lesssim 0.4$.

\subsection{Hopping amplitude between subsystem and center sites}

\begin{figure}[t] 
\begin{center}
\includegraphics[clip,width=6.5cm]{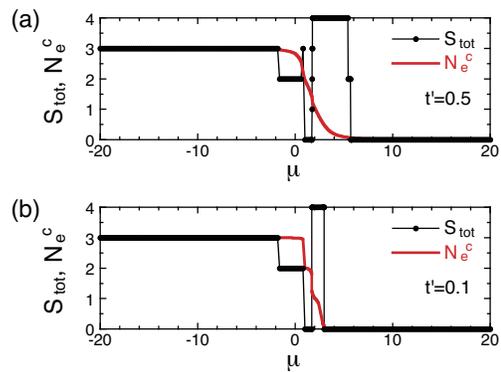}
\end{center}
\caption{(Color online)
The total spin $S_{\rm tot}$ and
the number of electrons in the center sites $N_{\rm e}^{\rm c}$ for
(a) $t'=0.5$ and
(b) $t'=0.1$
at $U=1000$ for $N=11$.
Note that the plot for $t'=1.0$ was given in Fig.~\ref{Fig_stot-nc-U1000}(a).
}
\label{Fig_stot-nc_tp}
\end{figure}

\begin{figure}[t] 
\begin{center}
\includegraphics[clip,width=6.5cm]{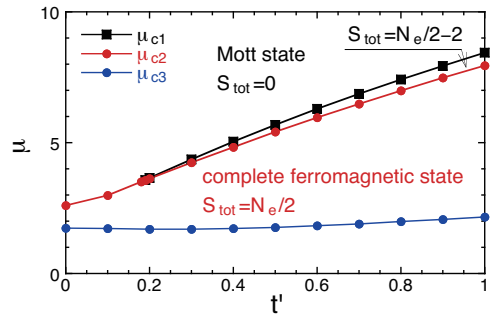}
\end{center}
\caption{(Color online)
The ground-state phase diagram in the coordinate $(t',\mu)$
at $U=1000$ for $N=11$.
The transition points $\mu_{\rm c}$'s are defined
in the same way as those in Fig.~\ref{Fig_DMRG}.
}
\label{Fig_muc-tp_n11ne8}
\end{figure}

Now, let us introduce a different value of the hopping amplitude
between the subsystem and the center sites $t'$
from that within the subsystem $t$,
in order to control the degree of correlation
between the subsystem and the center sites in a direct way.
That is,
the center sites are separated from the subsystem in the limit of $t'=0$.
In Fig.~\ref{Fig_stot-nc_tp}
we show the $\mu$ dependence of $S_{\rm tot}$ and $N_{\rm e}^{\rm c}$
for $t'=0.5$ and $0.1$.
The plot for $t'=1$ was already given in Fig.~\ref{Fig_stot-nc-U1000}(a).
As $t'$ decreases,
the region where $N_{\rm e}^{\rm c}$ increases from zero to three ($=N_{\rm c}$)
becomes narrower,
while $N_{\rm e}^{\rm c}$ exhibits a sharp change with a plateaulike structure at integers.
There appears a compelete ferromagnetic state in a region where
$N_{\rm e}^{\rm c}$ varies from zero to one.

In Fig.~\ref{Fig_muc-tp_n11ne8}
we depict the ground-state phase diagram in the coordinate $(t',\mu)$ for $N=11$,
where $\mu_{\rm c1}$, $\mu_{\rm c2}$, and $\mu_{\rm c3}$ are defined in the same way
as those in Fig.~\ref{Fig_muc-U_n11ne8}.
We observe that
$\mu_{\rm c3}$ does not depend on $t'$ so much,
while $\mu_{\rm c2}$ is reduced as $t'$ decreases.
As a result,
the range of $\mu$ of the complete ferromagnetic state becomes narrow with decreasing $t'$.

In the limit of $t'=0$,
the subsystem and the center sites are independent by definition.
We find that in the period $\mu_{\rm c2} < \mu < \mu_{\rm c3}$,
the subsystem contains just seven electrons
and the remaining one electron stays at a center site.
In this situation,
the Nagaoka ferromagnetism is realized in the subsystem, i.e., $S_{\rm sub}=7/2$,
while $S_{\rm c}=1/2$.
Here the states of $S_{\rm tot}=S_{\rm sub}\pm S_{\rm c}$ are degenerate,
and thus we obtain the complete ferromagnetic state.
Note that in this limit
we do not have the phase of $S_{\rm tot}=N_{\rm e}/2-2$.
We find that $\mu_{\rm c1}$ appears above $t'\simeq 0.19$.

\subsection{Entanglement entropy}

In Figs.~\ref{Fig_CSXY} and \ref{Fig_CSZ}
we found that the spin correlation
between spins in the subsystem and those in the center sites are very weak
regardless of the situation of the total spin.
Note that the spin correlation is still weak
even in the case of $\mu=0$ (not shown).
However, we expect that
strong quantum correlations occur in the ferromagnetic state
where all the spins contribute to form the totally symmetric wave function.

Now we study the nature of correlation in the system
by making use of the entanglement entropy
\cite{Amico2008}
instead of the spin correlation functions.
In particular, we measure the entanglement entropy of the subsystem
by tracing out the degrees of freedom in the center sites.
The reduced density matrix of the subsystem is given by
\begin{equation}
  \rho_{\rm sub}={\rm Tr}_{\rm c}|G\rangle\langle G|,
\end{equation}
where $|G\rangle$ is the ground state of the whole system,
and the degrees of freedom of the center sites are traced out.
The entanglement entropy of the subsystem is defined by
\begin{equation}
  E_{\rm sub}=-{\rm Tr}_{\rm sub}\rho_{\rm sub}\ln\rho_{\rm sub},
\end{equation}
where the trace is taken for the subsystem.
We note that the entanglement entropy of the center sites can also be defined as
$E_{\rm c}=-{\rm Tr}_{\rm c}\rho_{\rm c}\ln\rho_{\rm c}$
with
$\rho_{\rm c}={\rm Tr}_{\rm sub}|G\rangle\langle G|$.
The relation $E_{\rm sub}=E_{\rm c}$ holds for the ground state,
although, in general, $E_{\rm sub} \neq E_{\rm c}$ for a mixed state.

\begin{figure}[t] 
\begin{center}
\includegraphics[clip,width=8cm]{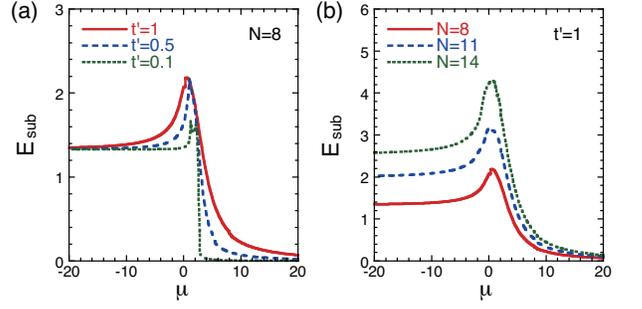}
\end{center}
\caption{(Color online)
(a) The entanglement entropy of the subsystem $E_{\rm sub}$ for several values of $t'$ for $N=8$.
(b) $E_{\rm sub}$ at $t'=1$ for several system sizes.
Here $U=1000$.
}
\label{Fig_EE}
\end{figure}

In Fig.~\ref{Fig_EE}(a)
we present the $\mu$ dependence of $E_{\rm sub}$ for several values of $t'$.
In the Mott-state regime at large $\mu$,
since the center sites are empty,
the subsystem and the center sites are practically separated.
Therefore, $E_{\rm sub}$ is suppressed,
and it approaches zero in the limit of $\mu\rightarrow\infty$.
As $t'$ decreases,
$E_{\rm sub}$ decays to zero at lower $\mu$,
since the subsystem and the center sites are disconnected.

With decreasing $\mu$,
electrons turn to come to the center sites as well as the subsystem,
and thus we expect that
the correlation between the subsystem and the center sites becomes significant.
In the region of the complete ferromagnetic state,
the fully symmetrized state of all the spins takes place,
and $E_{\rm sub}$ is enhanced, as expected.
We find a peak near $\mu=0$
where electrons can move around the whole system
without the disturbance of the chemical potential.
This enhancement of the entanglement is due to the electron motion,
i.e., the subsystem is entangled by the quantum motion
among the subsystem and the center sites.
We note that the entanglement occurs
even in the limit of $t'=0$.

Here let us discuss the behavior for large negative $\mu$,
where we find that
$E_{\rm sub}$ converges to a constant value as $\mu$ decreases independent of $t'$.
We can understand this convergent behavior
by considering the contribution of relevant spin states as follows.
For large negative $\mu$, each center site traps one electron.
Note that the contribution of the double occupied states is negligibly small
because of the Coulomb repulsion unless we do not decrease $\mu$ down to $-U$.
If $t'$ is small,
the center sites are isolated from the system,
so that we can consider subspaces
with different spin states in the center sites independently.
In the case with $N=8$ and $N_{\rm e}=6$,
we have two electrons in the two center sites,
and there appear four spin states,
i.e., $(\uparrow,\uparrow)$, $(\downarrow,\downarrow)$,
$(\uparrow,\downarrow)$, and $(\downarrow,\uparrow)$.
These four states correspond to the subspaces with
$(N_{\uparrow}^{\rm sub},N_{\downarrow}^{\rm sub})=(1,3)$,
$(3,1)$, $(2,2)$, and $(2,2)$, respectively,
where $N_{\sigma}^{\rm sub}$ denotes
the number of spin-$\sigma$ electrons in the subsystem.
The electron hopping $t'$ mixes the states of the lowest energies of the four cases,
which gives the entanglements.
That is, the exchange of the spin state between the subsystem and the center sites
brings the entanglements.
As the first approximation,
let us assume that the four states have an equal weight in the ground state,
i.e., the four states are maximally entangled.
Then, the entanglement entropy is given by
\begin{equation}
  \tilde{E}_{\rm sub} =
  -\sum_{(\uparrow,\uparrow),(\downarrow,\downarrow),(\uparrow,\downarrow),(\downarrow,\uparrow)}
  \frac{1}{4}\ln \frac{1}{4} = 1.386,
\end{equation}
where the summation is taken over allowed patterns of
the spins trapped in the center sites.
The thus obtained value is close to the result in Fig.~\ref{Fig_EE}(a),
i.e., $E_{\rm sub} \simeq 1.328$ at $\mu=-20$ for $t'=0.1$.
Here we find a slight difference,
because relative weights of the four states are different in reality.
If we take account of the relative weights,
we have a more precise estimation of $E_{\rm sub}$.
In fact, the four largest eigenvalues of $\rho_{\rm sub}$ are
$\lambda(\uparrow,\uparrow)=\lambda(\downarrow,\downarrow)\simeq 0.167$,
$\lambda((\uparrow,\downarrow)+(\downarrow,\uparrow))\simeq 0.310$,
and
$\lambda((\uparrow,\downarrow)-(\downarrow,\uparrow))\simeq 0.357$,
and these eigenvalues gives the entanglement entropy
\begin{eqnarray}
  \tilde{E}_{\rm sub}
  &=& -\sum_{i=1}^{4} \lambda_{i}\ln\lambda_{i}
\nonumber\\
  &=& -0.167 \times \ln 0.167
      -0.167 \times \ln 0.167
\nonumber\\
  &&  -0.310 \times \ln 0.310
      -0.357 \times \ln 0.357
\nonumber\\
  &=& 1.328,
\end{eqnarray}
which gives a good agreement with the numerical result.
We notice that the relative weights of the four states do not change so much
even when $t'$ is varied,
although they are not defined at $t'=0$.

In Fig.~\ref{Fig_EE}(b)
we depict the size dependence of $E_{\rm sub}$ at $t'=1$.
We see that the $\mu$ dependence is the same in shape for different system sizes,
while $E_{\rm sub}$ becomes large in the whole range of $\mu$
with increasing $N$.
In particular, 
$E_{\rm sub}$ converges to an $N$-dependent constant value
for large negative $\mu$.
Based on the same logic discussed for $N=8$ above,
as the first approximation,
the saturated value is described by the uniform distribution
of the eigenstates among allowed states,
\begin{equation}
  \tilde{E}_{\rm sub}
  =-\sum_{\rm allowed \ patterns}^{D}
   \frac{1}{D} \ln \frac{1}{D}
  =\ln D,
\end{equation}
where $D$ is the number of allowed patterns of the spins trapped in the center sites.
In the cases of $N=8$ and $11$,
$D=2^{N^{\rm c}}$,
since all the possible patterns are degenerate in energy.
For $N=11$
we estimate as $\tilde{E}_{\rm sub} = \ln 8 = 2.079$,
which is close to the result in Fig.~\ref{Fig_EE}(b),
i.e., $E_{\rm sub} \simeq 2.026$ at $\mu=-20$.
In contrast, in the case of $N=14$,
among the possible $2^4=16$ patterns,
we have three types of spin states:
(1) all up spins $(\uparrow\uparrow\uparrow\uparrow)$ and
all down spins $(\downarrow\downarrow\downarrow\downarrow)$;
(2) three up and one down spins $(\uparrow\uparrow\uparrow\downarrow)$ and
one up and three down spins $(\downarrow\downarrow\downarrow\uparrow)$;
and
(3) two up and two down spins $(\uparrow\uparrow\downarrow\downarrow)$.
We find that
$(\uparrow\uparrow\uparrow\uparrow)$ and $(\downarrow\downarrow\downarrow\downarrow)$
have a high energy compared to the others,
and $(\uparrow\uparrow\uparrow\downarrow)$ and $(\downarrow\downarrow\downarrow\uparrow)$
have an energy slightly higher than that of $(\uparrow\uparrow\downarrow\downarrow)$.
Thus we exclude the first case and we obtain $D=14$.
Then we estimate as $\tilde{E}_{\rm sub}=\ln 14 = 2.639$,
which is again close to the result in Fig.~\ref{Fig_EE}(b),
i.e., $E_{\rm sub} \simeq 2.578$ at $\mu=-20$.
We note again that
the relative weights of the allowed patterns are different in reality,
so that measured values $E_{\rm sub}$ are slightly smaller than
estimations $\tilde{E}_{\rm sub}$.

\section{SUMMARY AND DISCUSSION}

In this paper we investigated the ferromagnetism in the itinerant electron model
by using numerical methods,
motivated by the Nagaoka ferromagnetism.
We pointed out that the magnetic property can be controlled with the non-magnetic parameter
such as the chemical potential representing the electron affinity.

As a concrete model
we adopted the Hubbard model (\ref{Eq_hammu}) which consists of two parts,
i.e., the subsystem and the center sites.
We examined how the total spin depends on the chemical potential for the center sites $\mu$,
which controls the distribution of the electron density.
We set the number of electrons
to be the same as the number of sites of the subsystem
to investigate the realization of the ``extended Nagaoka ferromagnetism.''
In the previous report
we have studied the minimal case of this type of lattice with five sites
to clarify dynamical aspects of the Nagaoka ferromagnetisim.~\cite{PTPmiya}
In the present paper
we extended the lattice to larger sizes as depicted in Fig.~\ref{Fig_extended-lattice},
and studied the case of hole doping into the subsystem at finite density.

Using Lanczos and DMRG methods
we examined how the itinerant ferromagnetism appears
as a function of $\mu$.
For large $\mu$,
the center sites are vacant and the subsystem is half-filled,
so that the ground state is the Mott state with zero total spin.
As $\mu$ decreases,
electrons can move from the subsystem to the center sites.
In other words, holes are doped into the subsystem.
We observed that the complete ferromagnetic state appears
in some range of $\mu$,
where the amount of holes doped into the subsystem is moderately small.

In the region of the complete ferromagnetic state,
we found that the total spin and the electron density exhibit
similar $\mu$ dependencies
for various values of the number of sites $N$,
suggesting a convergence when $N$ increases.
In particular,
the range of $\mu$ of the complete ferromagnetic state saturates well
at large values of $N$.
These results are indicative that
the present mechanism of the itinerant ferromagnetism exisits
with finite denisity of hole doping
in the thermodynamic limit. 
We discussed the nature of the ferromagnetism
by measuring various physical quantities
such as the spin correlation function, the magnetization process,
and the entanglement entropy.

We expect that
the present mechanism for the itinerant ferromagnetism can be realized
in some molecular-based magnets
in which highly mobile electrons exist
and the electron affinity at sites is modified by external forces.
Moreover, if the present setup of the Hubbard model with the local on-site energy
is prepared in the optical lattice of cold atoms in a controllable way,
the manipulation of the total spin can be performed.
We hope that our work stimulates the experimental exploration of
the itinerant ferromagnetism
by a new type of control of the magnetic property
through a lattice change, etc.

Finally, we note that
the effective electron number can be controlled by
the chemical potential at local sites in the whole system.
In the present system,
with the change of the chemical potential,
the number of electrons in the subsystem takes non-integers and changes continuously,
although the total number of electrons is an fixed integer.
This situation can be regarded as an emulation of the doping effect in the subsystem.
So far the doping effect has been studied
by changing the number of electrons one by one,
and the present scheme would provide an approach to study the doping effect.
We hope that this type of analysis 
can be applied to other systems with doping-induced phenomena,
such as superconductivity in doped systems,
which is an interesting future problem.

\section*{ACKNOWLEDGMENTS}

We are grateful to M. Okumura and M. Machida
for fruitful discussions.
The present work was supported by
Grant-in-Aid for Scientific Research (C) No.~25400391,
and the Elements Strategy Initiative Center for Magnetic Materials
under the outsourcing project of MEXT.
The numerical calculations were supported
by supercomputer centers of Japan Atomic Energy Agency
and ISSP of the University of Tokyo.



\begin{thebibliography}{99}

\bibitem{Hubbard1963}
J. Hubbard,
Proc. Roy. Soc. London Ser. A \textbf{276}, 238 (1963).

\bibitem{Kanamori1963}
J. Kanamori,
Prog. Theor. Phys. \textbf{30}, 275 (1963).

\bibitem{Gutzwiller1963}
M. C. Gutzwiller,
Phys. Rev. Lett. \textbf{10}, 159 (1963).

\bibitem{Stoner1946}
E. C. Stoner,
Rep. Prog. Phys. \textbf{11}, 43 (1947).

\bibitem{Nagaoka1966}
Y. Nagaoka,
Phys. Rev. \textbf{147}, 392 (1966).

\bibitem{Mielke1993}
A. Mielke and H. Tasaki,
Commun. Math. Phys. \textbf{158}, 341 (1993). 

\bibitem{Tasaki1998}
H. Tasaki,
Prog. Theor. Phys. \textbf{99}, 489 (1998).

\bibitem{Takahashi1982}
M. Takahashi,
J. Phys. Soc. Jpn. \textbf{51}, 3475 (1982).

\bibitem{Doucot1989}
B. Doucot and X. G. Wen,
Phys. Rev. B \textbf{40}, 2719 (1989).

\bibitem{Riera1989}
J. A. Riera and A. P. Young,
Phys. Rev. B \textbf{40}, 5285 (1989).

\bibitem{Fang1989}
Y. Fang, A. E. Ruckenstein, E. Dagotto, and S. Schmitt-Rink,
Phys. Rev. B \textbf{40}, 7406 (1989).

\bibitem{Shastry1990}
B. S. Shastry, H. R. Krishnamurthy, and P. W. Anderson,
Phys. Rev. B \textbf{41}, 2375 (1990).

\bibitem{Doucot1990}
B. Doucot and R. Rammal,
Phys. Rev. B \textbf{41}, 9617 (1990).

\bibitem{Putikka1992}
W. O. Putikka, M. U. Luchini, and M. Ogata,
Phys. Rev. Lett. \textbf{69}, 2288 (1992).

\bibitem{Becca2001}
F. Becca and S. Sorella,
Phys. Rev. Lett. \textbf{86}, 3396 (2001).

\bibitem{Carleo2011}
G. Carleo, S. Moroni, F. Becca, and S. Baroni,
Phys. Rev. B \textbf{83}, 060411 (2011).

\bibitem{Liu2012}
L. Liu, H. Yao, E. Berg, S. R. White, and S. A. Kivelson,
Phys. Rev. Lett. \textbf{108}, 126406 (2012).

\bibitem{Liang1995}
S. Liang and H. Pang,
Europhys. Lett. \textbf{32}, 173 (1995).

\bibitem{Kohno1997}
M. Kohno,
Phys. Rev. B \textbf{56}, 15015 (1997).

\bibitem{Watanabe1997}
Y. Watanabe and S. Miyashita,
J. Phys. Soc. Jpn. \textbf{66}, 2123 (1997).

\bibitem{Watanabe1999}
Y. Watanabe and S. Miyashita,
J. Phys. Soc. Jpn. \textbf{68}, 3086 (1999).

\bibitem{Kollar1996}
M. Kollar, R. Strack, and D. Vollhardt,
Phys. Rev. B \textbf{53}, 9225 (1996).

\bibitem{Stecher2010}
J. von Stecher, E. Demler, M. D. Lukin, and A. M. Rey,
New J. Phys. \textbf{12}, 055009 (2010).

\bibitem{Okumura2011}
M. Okumura, S. Yamada, M. Machida, and H. Aoki,
Phys. Rev. A \textbf{83}, 031606 (2011).

\bibitem{Lieb-Mattis}
E. Lieb and D. Mattis,
J. Math. Phys. \textbf{3}, 749 (1962).

\bibitem{PTPmiya}
S. Miyashita,
Prog. Theor. Phys. \textbf{120}, 785 (2008).

\bibitem{PRBmiya}
S. Miyashita, M. Ogata, and H. De Raedt,
Phys. Rev. B \textbf{80}, 174422 (2009).

\bibitem{White1992}
S. R. White,
Phys. Rev. Lett. \textbf{69}, 2863 (1992).

\bibitem{White1993}
S. R. White,
Phys. Rev. B \textbf{48}, 10345 (1993).

\bibitem{Rice1985}
T. M. Rice and K. Ueda,
Phys. Rev. Lett. \textbf{55}, 995 (1985).

\bibitem{Rice1986}
T. M. Rice and K. Ueda,
Phys. Rev. B \textbf{34}, 6420 (1986).

\bibitem{Guerrero1996}
M. Guerrero and R. M. Noack,
Phys. Rev. B \textbf{53}, 3707 (1996).

\bibitem{Batista2002}
C. D. Batista, J. Bon\v{c}a, and J. E. Gubernatis,
Phys. Rev. Lett. \textbf{88}, 187203 (2002).

\bibitem{Batista2003}
C. D. Batista, J. Bon\v{c}a, and J. E. Gubernatis,
Phys. Rev. B \textbf{68}, 214430 (2003).

\bibitem{Tsunetsugu1997}
H. Tsunetsugu, M. Sigrist, and K. Ueda,
Rev. Mod. Phys. \textbf{69}, 809 (1997).

\bibitem{Amico2008}
L. Amico, R. Fazio, A. Osterloh, and V. Vedral,
Rev. Mod. Phys. \textbf{80}, 517 (2008).

\end{thebibliography}
\end{document}